\documentclass{article}
\usepackage{amsmath,amsfonts,amssymb,graphicx,epsfig}
\usepackage{eurosym}
\usepackage{psfrag}
\usepackage{palatino}
\usepackage{colordvi,color}
\usepackage[normal]{subfigure}
\begin{document}

\title{Phase diagram of the asymmetric tetrahedral Ising-Heisenberg chain}
\author{J.S. Valverde, Onofre Rojas and S. M. de Souza \vspace{0.25cm} \\
%EndAName
{\small Departamento de Ci\^encias Exatas, Universidade Federal de Lavras} \\
{\small Caixa Postal 3037, CEP: 37200-000, Lavras, MG, Brazil }}
\maketitle

\begin{abstract}

The asymmetric tetrahedron is composed by all edges of tetrahedron represented by Ising interaction except one, which has a Heisenberg type interaction. This asymmetric tetrahedron is arranged connecting a vertex which edges are only Ising type interaction to another vertex with same structure of another tetrahedron. The process is replicated and this kind of lattice we call the asymmetric Ising-Heisenberg chain. We have studied the ground state phase diagram for this kind of models. Particularly we consider two situations in the Heisenberg-type interaction, (i) The asymmetric tetrahedral spin(1/2,1/2) Ising-XYZ chain, and (ii) the asymmetric tetrahedral spin-(1/2,1) Ising-XXZ chain, where we have found a rich phase diagram and a number of multicritical points. Additionally we have also studied their thermodynamics properties and the correlation function, using the decorated transformation. We have mapped the asymmetric tetrahedral Ising-Heisenberg chain in an effective Ising chain, and we have also concluded that it is possible to evaluate the  partition function including a longitudinal  external magnetic field.
\end{abstract}

\qquad %PACS numbers: 02.50.-r, 05.50.+q, 05.30.Fk\nonumber\\

\section{Introduction}

Low dimensional systems based on magnetic material have attracted considerable attention lately in a number of subjects such as condensed matter physics, material science and inorganic chemistry. In these particular areas  quantum ferrimagnetic chains (QFC) were discussed, due to that they exhibit a relevant combination of ferromagnetic (F) and antiferromagnetic (AF) states. Experimental synthetization of the compound Cu(3-Clpy)$_2$(N$_3$)$_2$\cite{escuer} with Clpy indicating Chloropyridine had been investigated. This compound could be mapped into a spin-1/2 tetramer chain with F-F-AF-AF bond alternation\cite{hagiwara}. Recently diamond type chain  structures have been intensively investigated theoretically and experimentally\cite{okamoto}. The natural candidates to describes these kinds of materials are the quantum anisotropic Heisenberg model or even Ising type models.  Certainly the rigorous mapping of those compounds into Heisenberg type models could become very complex systems which usually have non exact solution. However some particular case of models could become exactly solvable such as the Ising-Heisenberg model considered by Jascur and Strecka\cite{Jascur}, a more detailed discussion also  considered by Canova {et al.}\cite{Canova}.  The method used to solve this kind of models is the historical Fisher's work\cite{Fisher} of decorated transformation method, proposed  in 50's decade, the improvement of the method is discussed in reference \cite{Domb}. Many other  quasi-unidimensional Ising type models were solved using this methods\cite{Santos,Santos1}. 

Recently, theoretical investigation of strongly geometrical frustration materials\cite{Greedan}
have been performed, particularly focused on the diamond chain
structure, using several numerical approaches\cite{chen,Pati}. These
theoretical results could enhance the other experimental realizations
provided by the polymeric compounds such as Cu2OSO4\cite{belaiche} and M3(OH)2
(with M=Ni, Co, Mn)\cite{Guillou,wood}.
 Other quasi-unidimensional Heisenberg models were studied using numerical results such as discussed in references \cite{Pati,Niggemann} and some analytical series expansion also has been performed\cite{trh-prb} for similar systems.

This work aim is to present the frustrated properties of the asymmetric tetrahedron Ising Heisenberg (ATIH) chain, this models can be solved exactly mapping for an effective Ising chain with spins 1/2 or 1 using the method presented by Fisher\cite{Fisher}.  This work is organized as follow, in section 2, we present the ATIH chain, considering Heisenberg interaction with spin 1/2 and 1. In section 3, we discuss the phase diagram at zero temperature shown a rich phase diagram and several critical points, for the spin-(1/2,1/2) Ising-XYZ chain and spin-(1,1/2) Ising-XXZ chain.  In section 4, we discuss the thermodynamics properties for Ising-Heisenberg chain with Ising spin $s$=1/2 or 1 and Heisenberg interaction spin $S$=1/2 or 1, we also considered the correlation function using the decorated transformation method\cite{Fisher}. Finally in section 5, we present our conclusions. 

\section{The model}

The asymmetric tetrahedral Ising-Heisenberg (ATIH) chain is composed by all edges of tetrahedron (dashed line Fig.1) represented by Ising type interaction except one which is represented by Heisenberg type interaction (solid line in Fig. 1), which can be viewed also as Ising-Heisenberg diamond chain. To obtain the ATIH chain we coupled the vertex composed only by the Ising type {\it interaction edge}, we call from now on just as Ising {\it interaction vertex}, and connecting to another Ising {\it interaction vertex} of other tetrahedron. On the other hand the asymmetric edge of the tetrahedron are represented by Heisenberg type interaction, that we simply call as Heisenberg {\it interaction edge} (interaction between sites $a,b$ in Fig.1).   
\begin{figure}[h]
\begin{center}
\psfrag{Sc}{$s_c$}
\psfrag{S'a}{$S_a$}
\psfrag{S'b}{$S_b$}
\psfrag{J}{$J$}
\psfrag{Jab}{$J_{ab}$}
\includegraphics[width=11.cm,height=3.0cm,angle=0]{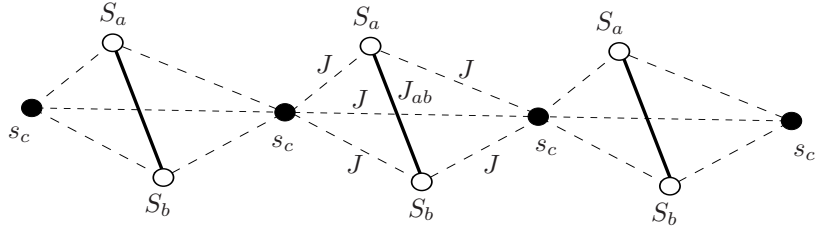}%
\end{center}
\caption[Diamond Ising chain]{The schematic representation of the coupled asymmetric tetrahedral Ising-Heisenberg. All edges (dashed line) of tetrahedral are represented by the Ising {\it interaction vertex} except one (solid line) which is represented by Heisenberg {\it interaction edge}.}
\end{figure}

The schematic representation of the ATIH model is given in Fig.1.  The Hamiltonian for the ATIH chain discussed above could be written by the following expression
\begin{equation} \label{a3}
H=\sum\limits_{i}H_{i,i+1}=\sum\limits_{i}\left[ J\left( S_{a,i}^{z}+S^{z}_{b,i}\right) \left( s_{c,i}+s_{c,i+1} \right) +Js_{c,i}s_{c,i+1}+ H_{i}^{XYZ}\right]. 
\end{equation}
with $s_{c,i}$ being the spin of Ising {\it interaction vertex}, where $J$ is the interaction parameter. 
The last term included in eq.\eqref{a3} correspond to the Heisenberg {\it interaction edge}, which is given by%
\begin{equation}\label{a3a}
H_{i}^{XYZ}=J_{x}S_{a,i}^{x}S_{b,i}^{x}+J_{y}S_{a,i}^{y}S_{b,i}^{y}+J_{z}S_{a,i}^{z}S_{b,i}^{z}, 
\end{equation}
being $S^{\nu}_i$ the spin matrices with $\nu=x,y,z$ and $J_{\nu}$ are their interaction parameter among sites $a$ and $b$.
We can also include the longitudinal external magnetic field in the Hamiltonian \eqref{a3}, which read as 
\begin{equation}
H_{m}=\sum\limits_{i}\left[ \frac{h_{0}}{2}\left( s_{c,i}+s_{c,i+1}\right)
+h\left( S_{a,i}^{z }+S_{b,i}^{z }\right) \right], 
\end{equation}%
here $h_{0}$ is an external magnetic field acting on spin $s_{c}$, whereas $h$ is an external magnetic field acting on $S_{a}^{z}$ and $S_{b}^{z}$. Note that we are considering different external magnetic fields because we assume that the gyromagnetic factor acting on $s_{c}$ could be different from that gyromagnetic factor acting on $S_{a}^{z}$ and $%
S_{b}^{z}$, which we report as $h_0=gh$ with $g$ being their relative gyromagnetic factor. The eq. \eqref{a3} is symmetric Hamiltonian in relation to the exchange $s_{c,i}\leftrightarrow s_{c,i+1}$ and $S_{a,i}^{z}\leftrightarrow S^{z}_{b,i}$. On the other hand we note that the Hamiltonian (\ref{a3}) also has an internal spin symmetry
$H( s_{c,i},s_{c,i+1}) =H(-s_{c,i},-s_{c,i+1}) \label{a3c}$.

\subsection{The XYZ {\it interaction edge} with spin-1/2}

To perform the partial summation over decorated site, we need to diagonalize the XYZ {\it interaction edge} Hamiltonian (\ref{a3a}), to evaluate this, we introduce the notations $J_{+}=J_{x}+J_{y}$ and $J_{-}=J_{x}-J_{y}$. For the spin $S=1/2$, we obtain the diagonalized Hamiltonian 
\begin{equation}
H_{i,i+1}=\text{diag}\left( \lambda _{+}^{(1)},\lambda _{+}^{(2)},\lambda
_{-}^{(2)},\lambda _{-}^{(1)}\right),  \label{a4}
\end{equation}%
with diag() we represents diagonal elements of the Hamiltonian \eqref{a3},
and conveniently we use the notation for simplicity $s_c=s_{c,i}$ and $s'_c=s_{c,i+1}$.
Thus the eigenvalues are given by%
\begin{align}
\lambda _{\pm }^{(1)} =&\gamma +\frac{1}{4}J_{z}\pm \frac{1}{4}\sqrt{16\alpha ^{2}+J_{-}^{2}},%
\text{ \ } \notag\\ \lambda _{\pm }^{(2)}=&\gamma -\frac{1}{4}J_{z}\pm \frac{1}{4}J_{+},  \label{a5} 
\end{align}
with
\begin{align}
\alpha\equiv \alpha(s_c,s'_c) =&J\left( {s_c}^{z}+{s'_c}^{z}\right) +h,\text{ \ } \label{alfa}\\ \gamma\equiv \gamma (s_c,s'_c)=&J{s_c}^{z}{s'_c}^{z}+\frac{h_{0}}{2}\left({s_c}^{z}+{s'_c}^{z}\right),
  \label{gama}
\end{align}%
here $\alpha(s_c,s'_c)$ and $\gamma (s_c,s'_c)$ are dependents of the spins $s_c$ and $s'_c$, thus we write from now on just as $\alpha$ and $\gamma$ respectively. Each eigenvalues are given by eq.\eqref{a5} which depend of $s_c$ and $s'_c$, then we have 16 eigenvalues.

 After  diagonalize the Hamiltonian function we get the corresponding set of eigenvectors also, thus the new basis are given by
\begin{align}
|v_{1}^{(+)}(s_c,s'_c)\rangle  =&\frac{1}{\sqrt{1+e_{1}^{2}}%
}( e_{1}|++\rangle +| --\rangle ) ,%
\text{ \ \ }&\quad |v_{1}^{(-)}(s_c,s'_c )\rangle =&\frac{1}{\sqrt{%
1+e_{2}^{2}}}( e_{2}\vert ++\rangle +\vert
--\rangle ),   \label{e1a} \\
|v_{2}^{\left( +\right) }\rangle =&\frac{1}{\sqrt{2}}\left( \left\vert
+-\right\rangle +\left\vert -+\right\rangle \right) ,\text{ \ \ }& \quad 
|v_{2}^{\left( -\right) }\rangle=&\frac{1}{\sqrt{2}}\left( -\left\vert
+-\right\rangle +\left\vert -+\right\rangle \right),   \label{e1b}
\end{align}%
where the factors $e_{1}$, $e_{2}$ depends of spins $s_c$ and $s'_c$ which are given by the following values
\begin{equation}
e_1\equiv e_{1}(s_c,s'_c)=\frac{\sqrt{16\alpha ^{2}+J_{-}^{2}}+4\alpha }{%
J_{-}},\quad \quad e_2\equiv e_{2}(s_c,s'_c) =\frac{-\sqrt{16\alpha
^{2}+J_{-}^{2}}+4\alpha }{J_{-}},  \label{e1c}
\end{equation}%
the normalized eigenvectors $|v_{1}^{( \pm ) }(s_c,s'_c)\rangle$ and $|v_{2}^{(\pm)}\rangle$
belonging to the eigenvalues $\lambda _{1}^{\left( \pm \right) }$ and $\lambda
_{2}^{\left( \pm \right) }$ given by \ (\ref{a5}). We also remark that the eigenvectors $|v_{2}^{\left( \pm \right) }\rangle$ are independents of the spins $s_c$ and $s'_c$ or we can say it is four-fold degenerated.  

At this point we would like to comment about the relations of the $e_{1}(s_c,s'_c)$ and $e_{2}(s_c,s'_c)$ factors. From (\ref{e1c}), it is not difficult to note that these factors transform one eigenvector of eq.\eqref{e1a} into the other one, when we exchange the following values,
\begin{eqnarray}\label{exchange}
\left\{
\begin{array}{l}
s_c\rightarrow-s_c\\
s'_c\rightarrow-s'_c\\
J_{-}\rightarrow-J_{-}\\
h\rightarrow-h
\end{array}\right\}
\Longrightarrow |v_{1}^{(+)}(s_c,s'_c)\rangle \rightarrow |v_{1}^{( -) }(s_c,s'_c)\rangle,
\end{eqnarray}
this property could be valid in more general situation, even in the presence of an external magnetic field.  However, it is important to point out that the eigenvalues of eq.(\ref{a5}) does not have a similar transformation. The eigenvalues $\lambda_{+}^{(1)}$ will always be at higher energy level than the eigenfunction $\lambda_{-}^{(1)}$. In the case of null magnetic field ($h=0,h_{0}=0$) we have these coefficients given by 
\begin{align}
e_{1}\left( \pm, \mp \right) &=-e_{2}\left( \pm, \mp \right)=1,&\\ %
e_{1}\left( \pm, \pm \right) &=\frac{\sqrt{16J^{2}+J_{-}^{2}}\pm 4J}{J_{-}},%
\quad& e_{2}\left( \pm, \pm \right) &=\frac{-\sqrt{16J^{2}+J_{-}^{2}}\pm
4J}{J_{-}}.  \label{e2}
\end{align}%

These properties will be useful when we discuss the phase diagram properties.
\subsection{The XXZ {\it interaction edge} with spin-1}

Now we can consider XXZ interaction among sites $a$ and $b$ (see fig.1),
with spin $S=1$ case.  We diagonalize the Hamiltonian analogous to the previous case. After diagonalizing the Hamiltonian depends only of $s_c$ and $s'_c$ and read as
\begin{equation}
H_{i,i+1}=\text{diag}\left( \lambda _{+}^{(1)},\lambda _{+}^{(2)},\lambda
_{+}^{(3)},\lambda _{+}^{(4)},\lambda ^{(5)},\lambda _{-}^{(4)},\lambda
_{-}^{(3)},\lambda _{-}^{(2)},\lambda _{-}^{(1)}\right),  \label{a7}
\end{equation}%
with diag() we represents diagonal elements of the Hamiltonian \eqref{a3},
whereas the eigenvalues are given by
\begin{align}\label{a8-1}
\lambda _{\pm }^{(1)} =&\pm 2\alpha +\gamma +J_{z}, \\ \lambda _{\pm
}^{(2)}=&\pm \alpha +\gamma +\frac{1}{2}J_{+},\text{ \ } \\ \lambda _{\pm
}^{(3)}=&\pm \alpha +\gamma -\frac{1}{2}J_{+},   \\
\lambda _{\pm }^{(4)} =&\gamma -\frac{1}{2}J_{z}\pm \frac{1}{2}\sqrt{%
J_{z}^{2}+2J_{+}^{2}},\text{ \ } \\ \lambda ^{(5)}=&\gamma -J_{z},  \label{a8}
\end{align}%
where  $\alpha$ and $\gamma$ already was defined by the eqs.\eqref{alfa} and \eqref{gama}.

Now for a complete analyze of the ATIH model, we shall turn our attention to study the $XXZ$ {\it interaction edge} for the decorated spin-1. In this situation we have nine eigenvector after diagonalizing the Hamiltonian, corresponding to the eigenvalues (\ref{a8-1}-\ref{a8}), thus the normalized eigenvectors are 
\begin{align}\label{sp1}
|u_{1}^{+}\rangle =&|1,1\rangle, &|u_{1}^{-}\rangle,  =&|-1,-1\rangle,&\\
|u_{2}^{+}\rangle=&\dfrac{1}{\sqrt{2}}\left( |1,0\rangle + |0,1\rangle\right),  & |u_{2}^{-}\rangle=&\dfrac{1}{\sqrt{2}} \left( |0,-1\rangle + |-1,0\rangle\right),& \\
|u_{3}^{+}\rangle =&\dfrac{1}{\sqrt{2}}\left( -|1,0\rangle + |0,1\rangle\right) ,& |u_{3}^{-}\rangle=&\dfrac{1}{\sqrt{2}}\left(-|0,-1\rangle + |-1,0\rangle\right),&\\
|u_{4}^{+}\rangle =&\dfrac{1}{\sqrt{2+f_{1}^{2}}}\left( |1,-1\rangle + |-1,1\rangle +f_{1}|0,0\rangle \right),&  |u_{4}^{-}\rangle=&\dfrac{1}{\sqrt{2+f_{2}^{2}}}\left( |1,-1\rangle + |-1,1\rangle +f_{2}|0,0\rangle \right),&\\
|u_{5}\rangle=&\dfrac{1}{\sqrt{2}}\left( -|1,-1\rangle + |-1,1\rangle\right),&&& \label{sp4}
\end{align}
where $f_1$ and $f_2$ are respectively given by
\begin{equation}
f_1=\dfrac{J_{z}+\sqrt{J_{z}^{2}+8J_{x}^{2}}}{2J_{x}}, \text{ \ \ }\quad f_2=\dfrac{J_{z}-\sqrt{J_{z}^{2}+8J_{x}^{2}}}{2J_{x}},
\end{equation}
at this point, we have a similar situation as was pointed out in eq.(\ref{exchange}), i.e., these factors transform as 
\begin{equation}
J_{z}\rightarrow -J_{z}, \text{ \ \ }J_{x}\rightarrow -J_{x} \quad \text{then}\quad |u_{4}^{+}\rangle \rightarrow |u_{4}^{-}\rangle, \label{sp4a}
\end{equation} 
again this conclusion is valid even when an external magnetic field is included.

After rewritten the Heisenberg {\it interaction edge} Hamiltonian in the diagonal form we are able to discuss the phase diagram for whole quasi-unidimensional chain for both cases XYZ {\it interaction edge} with spin-1/2 and XXZ {\it interaction edge} with spin-1.

\section{The Phase diagrams of the ATIH chain}

\subsection{The asymmetric tetrahedral spin-(1/2,1/2) Ising-XYZ chain}
To study the phase diagram of the asymmetric tetrahedral spin-(1/2,1/2) Ising-XYZ chain we use the diagonalized version of the Hamiltonian presented in the previous section 2.1.
We would like to note that for the present considered model, we have a set of sixteen different state vectors, nevertheless, when the translation and the global spin inversion symmetry are taking into account, we obtain that only eight state vectors have different phases.

 As we mentioned above some state vectors are physically equivalents, for example, the state vectors $| v_{1}^{(
+)}(+,+)\rangle $ and $|v_{1}^{(+)}(-,-)\rangle $ corresponding to the same state when we consider the global spin inversion. Once the eigenvalues satisfy the  relations $\lambda _{\pm}^{(1)}(s_c,s'_c)=\lambda _{\pm}^{(1)}(-s_c,-s'_c)$ and $\lambda_{\pm}^{(2)}(s_c,s'_c)=\lambda_{\pm}^{(2)}(-s_c,-s'_c)$, which restrict the eigenvectors to only eight relevant energy states.
 So we obtain the following eight states vectors for the asymmetric tetrahedral spin-(1/2,1/2) Ising-XYZ chain%
\begin{align}
\left|QFO_I\right\rangle  =&\prod\limits_{k=1}^{N}\left\vert +,v_{1}^{(+) }\left(+,+\right)\right\rangle _{k},& m_{0}=&0.5,\quad  m_1=\tfrac{1}{2}\big(\tfrac{e_1^2-1}{e_1^2+1}\big)&\label{e3} \\
|QFO_{II}\rangle =&\prod\limits_{k=1}^{N}\left\vert +, v_{1}^{\left( -\right) }\left(+,+\right) \right\rangle _{k},& m_{0}=&0.5, \quad  m_1=\tfrac{1}{2}\big(\tfrac{e_2^2-1}{e_2^2+1}\big)& \label{e4} \\
|QFO_{III}\rangle =&\prod\limits_{k=1}^{N}\left\vert +,v_{2}^{\left( +\right) }\right\rangle _{k},& m_{0}=&0.5,\quad m_{1}=0,&  \label{e4a} \\
|QFO_{IV}\rangle=&\prod\limits_{k=1}^{N}\left\vert +,v_{2}^{\left( -\right) }\right\rangle _{k},& m_{0}=&0.5,\quad m_{1}=0,&  \label{e4b} \\
|FRU_{I}\rangle =&\prod\limits_{k=1}^{N/2}\left\vert +,v_{1}^{\left( +\right) }\left( +,-\right),-,v_{1}^{\left( +\right) }\left(+,-\right) \right\rangle _{k},& m_{0}=&0,\quad m_{1}=0,&
\label{e5} \\
|FRU_{II}\rangle =&\prod\limits_{k=1}^{N/2}\left\vert +,v_{1}^{\left( -\right) }\left( +,-\right),-,v_{1}^{\left( -\right) }\left(+,-\right) \right\rangle _{k},& m_{0}=&0,\quad m_{1}=0,&
\label{e6} \\
|FRU_{III}\rangle =&\prod\limits_{k=1}^{N/2}\left\vert +,v_{2}^{\left( +\right) },-,v_{2}^{\left( +\right) }\right\rangle _{k},& m_{0}=&0,\quad m_{1}=0,&  \label{e7} \\
|FRU_{IV}\rangle=&\prod\limits_{k=1}^{N/2}\left\vert +,v_{2}^{\left( -\right)},-,v_{2}^{(-)}\right\rangle _{k},& m_{0}=&0,\quad  m_{1}=0,&  \label{e8}
\end{align}
where the first element inside the product corresponds to the Ising interaction taking two possible values ($\pm 1$), and the next element represent the XYZ {\it interaction edge} considered in the previous section.
All products are carried out over all spin sites.  In these relations the single Ising site magnetization $m_{0}$ is given for the spin-1/2 and $m_{1}$ is the single Heisenberg magnetization for the $a$, $b$ sites (Heisenberg {\it interaction edge}). The first two states (\ref{e3}) and (\ref{e4}) are new states and arise when we consider the total Ising-$XYZ$ case and their magnetization $m_1$ depends of the parameters $J$, $J_{-}$ and $h$. These states we will call in general "Quantum ferromagnetic" (QFO) states of type I and II, respectively, which are not degenerated and from (\ref{e1a}) it is possible to see that the probability of the spin alignment, defined by the functions $e_{1}(s_c,s'_c)$ and $e_{2}(s_c,s'_c)$ are not equivalents for the up and down orientation. Worth remark that this "Quantum ferromagnetic" state could become Quantum Ferrimagnetic state when $m_1$ or $m_2$ is negative. The energy states of (\ref{e4a}) and (\ref{e4b}) also have the structure of the first two states but now the probability of the up and down orientation are equivalent, these states are non-degenerated, we call also these state as QFO states of type III and IV respectively. The states (\ref{e5})-(\ref{e8}) display four frustrated (FRU) states of types I, II, III and IV respectively, and are non-degenerated. For these states we use an extended modified unitary cell necessary for the identification of equivalent vector states.

In Fig.2, we show the ground-state phase diagram for the system in the absence of the magnetic field ($h=h_0=0$). In Fig.2a let us to consider the following re-parametrization for the interaction parameters, without loosing any physical generality,
\begin{equation}
J=\text{sin}(x),\text{  } J_{z}=\text{sin}(y),\text{  } J_{+}=4\text{cos}(y),\text{  } J_{-}=4\text{cos}(x).\label{e10} 
\end{equation}

Here we represent the interaction parameter by means of two new parameters $x$ and $y$. For the first two states (eq. \ref{e3} and \ref{e4}) we obtain that the Heisenberg {\it interaction edge} magnetization takes the value, $m_{1}=\sin(x)/2$, thus for $x=0,\pi $ we have $m_{1}=0,$ and for $x=\pi /2$ we obtain $m_{1}=1/2$. 

Certainly, by using this new parameters we
restrict the values of interaction parameters as follows: $|J|\leqslant 1$,
$|J_z|\leqslant 1$ and $|J_{\pm}|\leqslant 4$. In this limited region we have competing
interaction parameters, leading to several ground state energies. Out
of this region there are no new phases.

It is possible to see that in this phase diagram, we have five tricritical points where converge three states and two four-critical points where converge four states. In the last case we have that these points are located in the vertical line $x=\pi$, for this position the value obtained from (\ref{e10}) is $J=0, \text{\\} J_{z}=0.8$, i.e., in this region the system has pure Heisenberg interaction for $a,b$ sites, and in general, this region is defined by the line where a continuous phase transitions occurs. We are able to calculate the other interaction parameters corresponding to these two critical points $(x,y)$, thus we have that for the points $(\pi,0.9273)$ and $(\pi,2.2143)$ the interaction parameters takes the values $J_{x}=-1.7$, $J_{y}=2.3$ and $J_{x}=-3.2$, $J_{y}=0.8$, respectively. In
principle we must plot in Fig.2a for the parameters $x$ and $y$ in the
interval $[0,2\pi]$. But in order to highlight the rich region in the
interval $[0,\pi]$, we considered it only up to $3\pi/2$ the $x$
parameter, since we have the continuation of the states $QFO_{III}$,
$QFO_{II}$ and $QFO_{IV}$. On the other hand, in the $y$ axis we have also
repeated states ($FRU_I$, $FRU_{II}$, $FRU_{IV}$ and $QFO_{II}$) which are also
present in the interval considered in Fig. 2a. We will use the same
considerations in the next figures.

In the region where the axis takes the values $x<\pi$, we have that the interaction parameter $J>0$, i.e. it is positive defined, and for the region $x>\pi$ we obtain a negative value for the parameter $J<0$. In general negative values for interaction parameters favors the anti-parallel alignment for the spin ordered system. Thus define the type of phase ordered state whatever is $FRU$ or $QFO$ sates (left and right side at the axis $x=\pi$ in the diagram). At this stage it is necessary to point out that the phase state $QFO_{II}$ appears in both sectors (grey sector of Fig.2a), basically this occurs by the fact that the constant $e_{2}(s_c,s'_c)$ is different from zero. As will be mentioned below, if the constant $e_{2}(s_c,s'_c)$ comes close to zero, the region $QFO_{II}$ of the left hand side disappears.

The other five tricritical points can be found by a simple substitution, for example the point $(\pi/2,\pi/2)$ is where the phase states $FRU_{III}$, $FRU_{IV}$ and $QFO_{II}$ coexist. Other tricritical points are $(0.9273,1.4679)$, $(0.9273,1.6735)$, $(2.2143,1.4679)$, $(2.2143,1.6735)$.

\begin{figure}[h]
\begin{center}
\psfrag{FRU-(I)}{$FRU_{I}$}
\subfigure[ref1][]{\includegraphics[width=7.0cm,height=8.7cm,angle=-90]{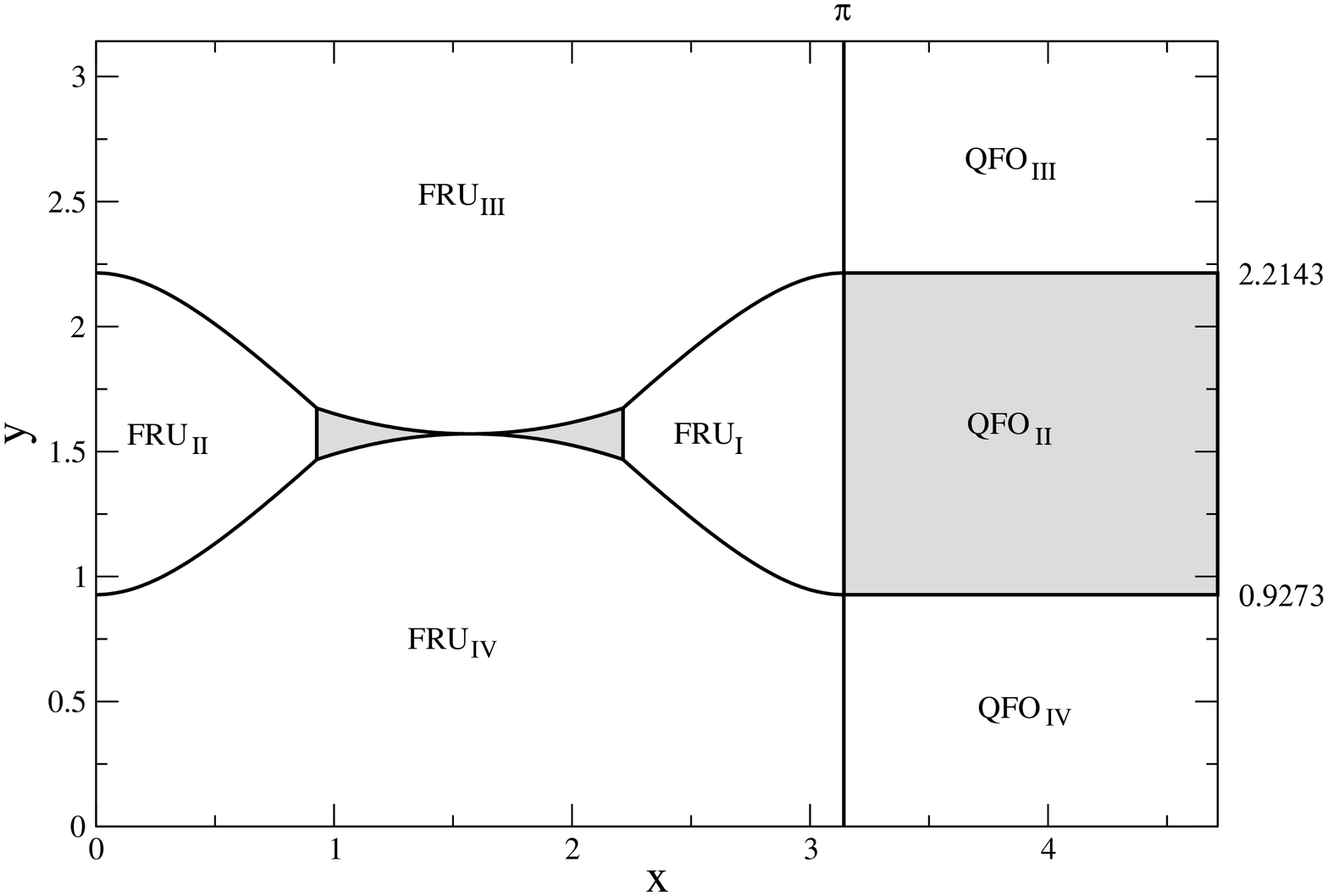}}
%\qquad
\subfigure[ref2][]{\includegraphics[width=7.0cm,height=8.7cm,angle=-90]{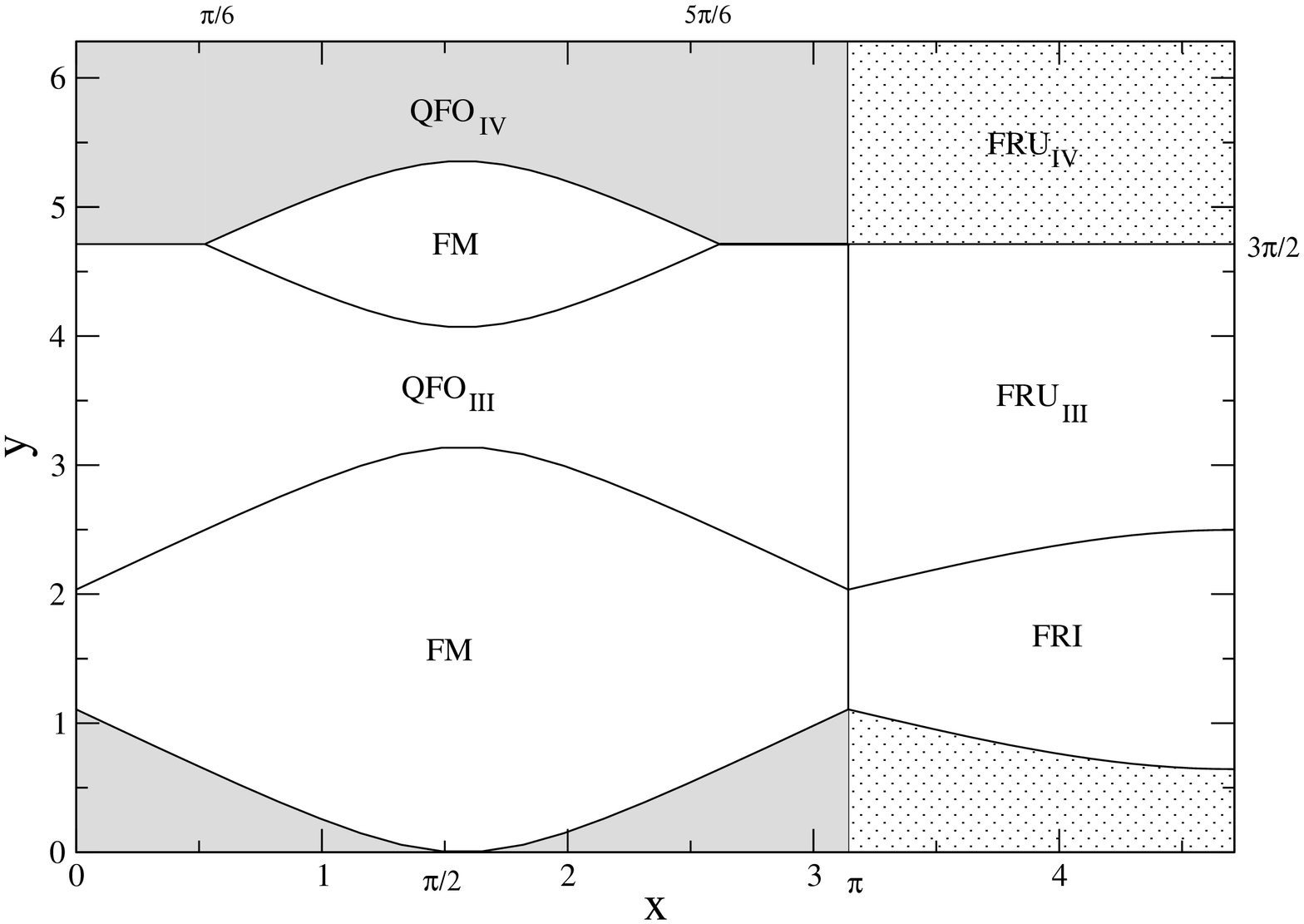}}
\end{center}
\caption{Schematic representation of the phase diagram. (a) For the  ATIH chain using the new parameters $x$ and $y$ defined by \eqref{e10}. (b) For the  ATIH chain using the new parameters $x$ and $y$ defined by \eqref{e10a}. }
\end{figure}

If we impose the condition $J_{x}=J_{y}$, i.e. for the $XXZ$ model, the eight states reduce to seven, six of them will appear in the phase diagram. This is possible because in the limit $(J_{-}\rightarrow 0)$ the constants $e_{1}(s_c,s'_c)$ and $e_{2}(s_c,s'_c)$ become $1$ and $0$ respectively, doing the states $FRU_{I}\rightarrow FRU_{II}$ equivalent.

We give in Fig.2(b), the ground-state energy for regions resulting from the substitution  $J_{-}=0$ and where six phases for the ground-states energy are shown. It is easy to see that the states vectors become $|v_{1}^{(+)}(s_c,s'_c)\rangle\rightarrow | ++\rangle$ and $|v_{1}^{(-)}(s_c,s'_c)\rangle\rightarrow | --\rangle $. For convenient reasons we use the following realization
\begin{equation}
J=-\text{sin}(x),\text{  } J_{x}=2\text{cos}(y),\text{  } J_{z}=-\text{sin}(y),\label{e10a} 
\end{equation}
and obtain now a ferromagnetic ($FM$) and ferrimagnetic ($FRI$) states. It is possible to see that the $x=\pi$ axis still present in these phase configuration. On this line three critical points are shown where four ground-states converge.
\begin{figure}[h]
\begin{center}
\subfigure[ref1][]{\includegraphics[width=7.0cm,height=8.7cm,angle=-90]{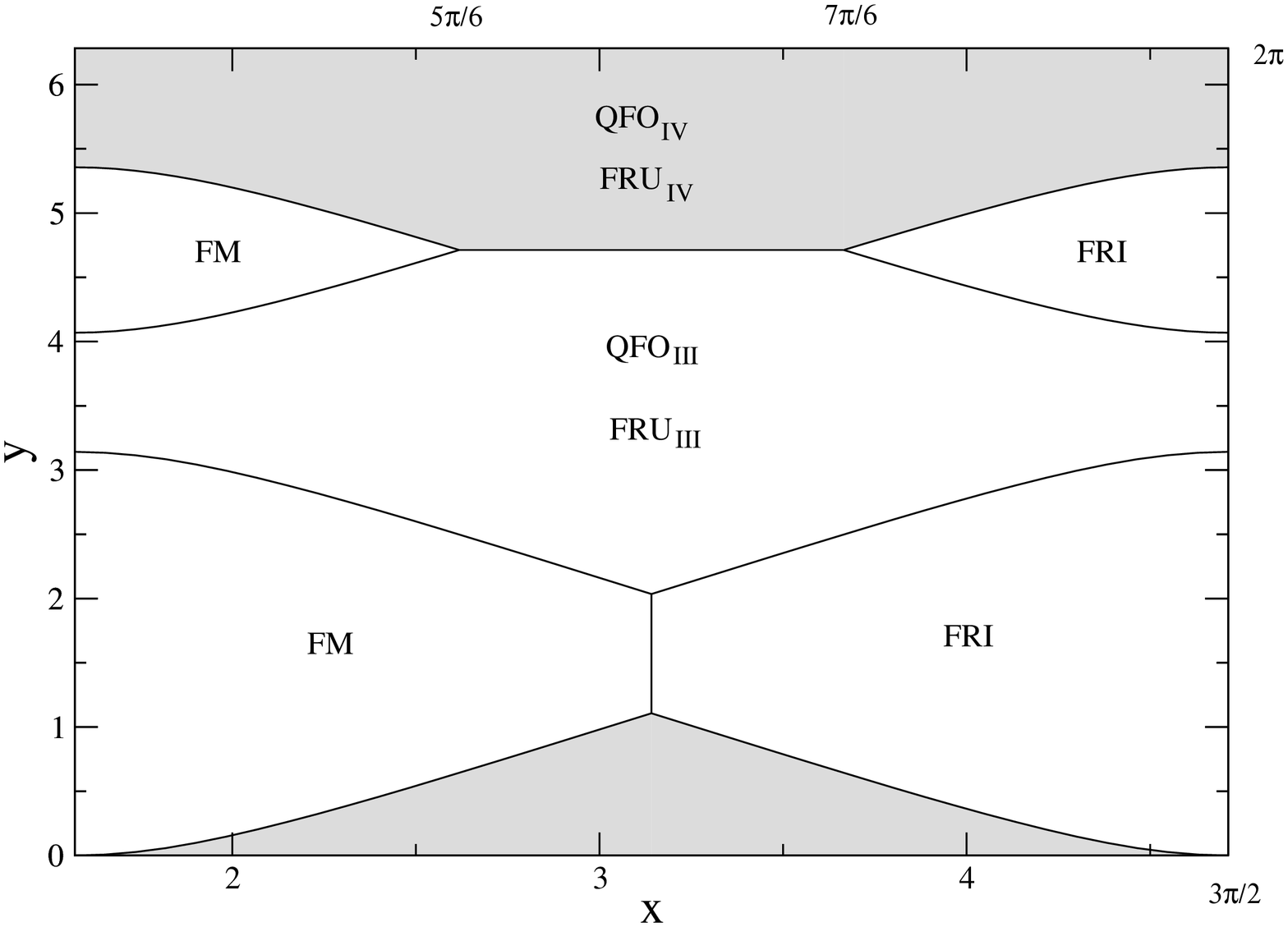}}
%\qquad
\subfigure[ref2][]{\includegraphics[width=7.0cm,height=8.7cm,angle=-90]{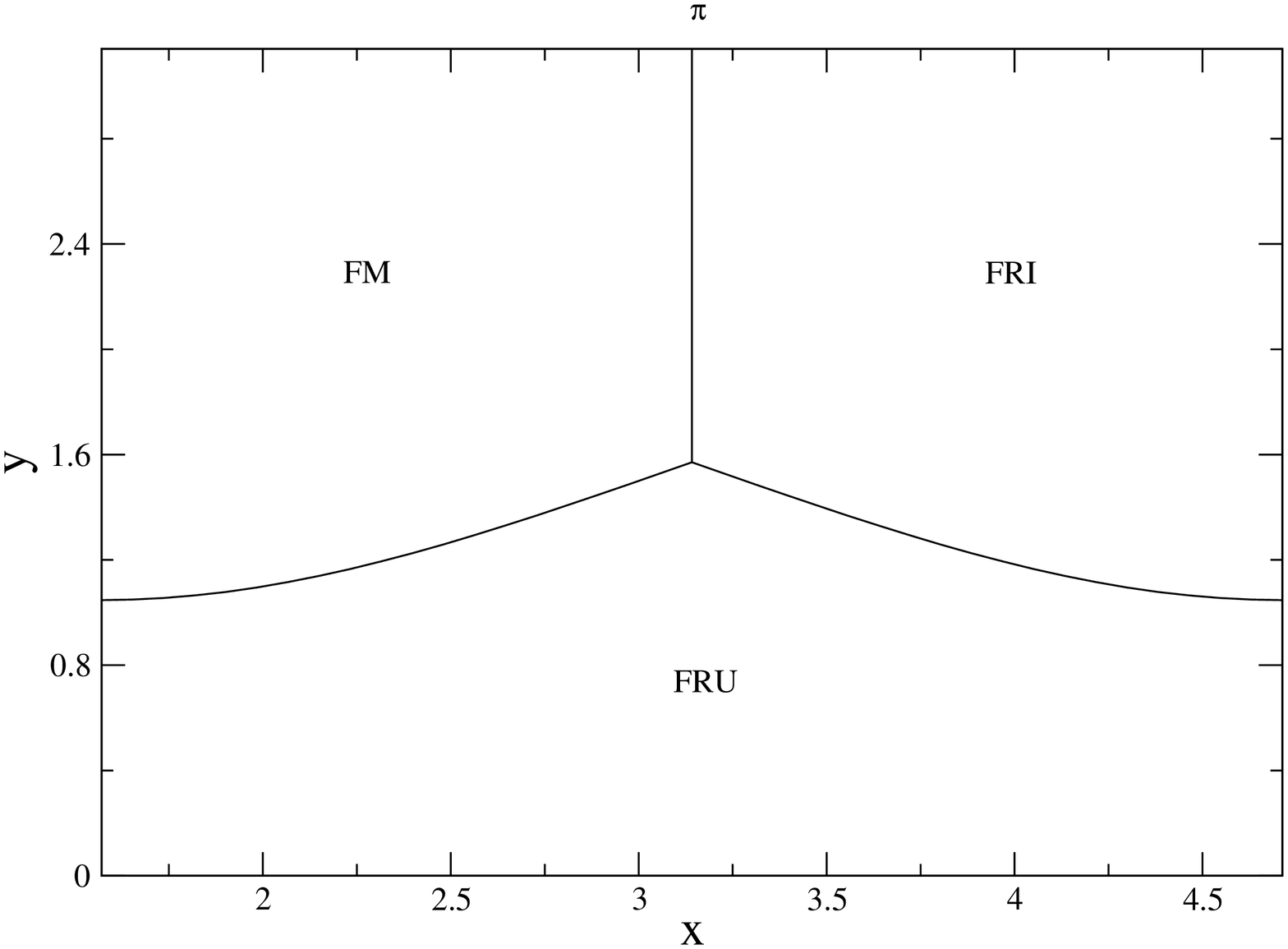}}
\end{center}
\caption{Schematic representation of the phase diagram, for ATIH  chain,  where we consider $\gamma=0$ and $J_{-}=0$. In \textit{(a)} we display a ferromagnetic $FM$ and ferrimagnetic $FRI$ states, however the quantum ferromagnetic $QFO$ and the frustrated $FRU$ phases are also present. In \textit{(b)}, we fixed the condition $J_{x}=J_{y}=J_{z}$ obtaining only three states phase energy. The $FM$, $FRI$ and $FRU$ sectors converges into one tricritical point $(\pi,\pi/2)$.}
\end{figure}
The phase diagram displayed in Fig.2, shows several states due to the presence of the  $Js_{c,i}s_{c,i+1}$ interaction, i.e. the $\gamma(s_c,s'_c)$ term (with $h_{0}=0$) given by (\ref{gama}) allows us to consider this interaction. In reference \cite{Canova} the term $Js_{c,i}s_{c,i+1}$  was considered null.  If we put $\gamma(s_c,s'_c)=0$ and at the same time impose the equality of some of the interaction parameters, we obtain a very simple phase diagram. This is graphically displayed in Fig.3. The Fig.3(a) shows that in the case $\gamma(s_c,s'_c)=0$, $J_{x}=J_{y}$, some states are degenerated, so we identify the $QFO_{IV}$ and $FRU_{IV}$ as having the same energy. The same occurs with the  $QFO_{III}$ and $FRU_{III}$ states. The $FM$ and $FRI$ states also appear in this diagram.
The Fig.3b shows that for the case $\gamma(s_c,s'_c)=0$, $J_{x}=J_{y}=J_{z}$ we obtain the more simple configuration with three phases, the $FM$, $FRI$ and $FRU$ states. Only one three-critical point where these states converges is present in $(\pi,\pi/2)$. At this stage we would like to remark that a very similar ground state configuration was obtained in \cite{Canova}, however, the realizations used in our work (\ref{e10}) and (\ref{e10a}) are different from those used in \cite{Canova}, because we did not include the external magnetic field.

\subsection{The asymmetric tetrahedral spin-(1/2,1) Ising-XXZ chain}

 To obtain all states we need to consider the coupling of the decorated vector states (\ref{sp1}-\ref{sp4}) with the Ising {\it interaction vertex} with spin-$1/2$, this enable us to write down the total vector states of the system. We will restrict the conditions over the interaction parameters without losing the generality, using the following values for the interaction parameters 
\begin{equation}
J=-J_{z}=\sin(x), \text{ \ \ } J_{x}=2\sin(y), \label{sp4b}
\end{equation}
in the following lines we also restrict the system to the case when the external magnetic field is absent ($h_{0}=0$, $h=0$). From all possible twenty-four ground-states energy, which can be obtained from the system only fifteen eigenvalues have different values. In this situation the ground-state eigenvectors for the asymmetric spin-(1/2,1) Ising-XXZ chain, that would appear in the phase diagrams are given by 
\begin{eqnarray}
|FM\rangle &=&\prod\limits_{k=1}^{N}\left\vert +,u_{1}^{\left( +\right) } \right\rangle _{k},\text{ \ \ } m_{0}=0.5,  \text{ \ \ } m_{1}=0.5, \label{sp5} \\
|FRI\rangle &=&\prod\limits_{k=1}^{N}\left\vert -,u_{1}^{\left( +\right) }\right\rangle _{k},\text{ \ \ }m_{0}=-0.5, \text{ \ \ } m_{1}=0.5, \label{sp6} \\
|QFO_I\rangle &=&\prod\limits_{k=1}^{N}\left\vert +,u_{2}^{\left( +\right) }\right\rangle
_{k},\text{ \ \ } m_{0}=0.5,\text{ \ \ } m_{1}=0.5,  \label{sp7} \\
|QFO_{II}\rangle &=&\prod\limits_{k=1}^{N}\left\vert +,u_{3}^{\left( +\right) }\right\rangle
_{k},\text{ \ \ } m_{0}=0.5,\text{ \ \ } m_{1}=0.5,  \label{sp8} \\
|QFO_{III}\rangle &=&\prod\limits_{k=1}^{N}\left\vert +,u_{4}^{\left( -\right) } \right\rangle _{k},\text{ \ \ } m_{0}=0.5,\text{ \ \ } m_{1}=0,
\label{sp9} \\
|QFI_{I}\rangle &=&\prod\limits_{k=1}^{N}\left\vert -,u_{2}^{\left( +\right) } \right\rangle _{k},\text{ \ \ } m_{0}=-0.5,\text{ \ \ } m_{1}=0.5,
\label{sp10} \\
|QFI_{II}\rangle &=&\prod\limits_{k=1}^{N}\left\vert -,u_{3}^{\left( +\right) }\right\rangle
_{k},\text{ \ \ } m_{0}=-0.5,\text{ \ \ } m_{1}=0.5,  \label{sp11} \\
|FRU\rangle &=&\prod\limits_{k=1}^{N/2}\left\vert +,u_{4}^{\left( -\right) },-u_{4}^{\left( -\right) }\right\rangle
_{k},\text{ \ \ } m_{0}=0\text{ \ \ } m_{1}=0,  \label{sp12}
\end{eqnarray}
where the first element inside the product corresponds to the Ising interaction taking two possible values ($\pm 1$), and the next element represents the XXZ {\it interaction edge} considered in the previous section 2.2.
 The eq.\eqref{sp5} represents a ferromagnetic (FM) state, eq.\eqref{sp6} indicate a ferrimagnetic (FRI) state, whereas eqs. \eqref{sp7}-\eqref{sp9} corresponds to Quantum ferromagnetic (QFO) state of type I, II and III respectively. We also have two types of quantum ferrimagnetic (QFI) states given by eq.\eqref{sp10} and \eqref{sp11}, and finally eq.\eqref{sp12} represents a frustrated (FRU) state.
All these states given by (\ref{sp5}-\ref{sp12}) are displayed in the phase diagram presented in Fig.4(a). It is remarkable that the line $x=\pi$ divides the $QFO$ states from the other ones. Therefore in this case we have a seven-critical point in ($\pi,\pi$) where all states converge. 

In order order to display the first vector state (\ref{sp5}), which not present in fig.4a,  we change the restriction over the interaction parameters and fixed as follow
\begin{equation}
J=J_{z}=\sin(x), \text{ \ \ } J_{x}=2\sin(y), \label{sp13}
\end{equation}
the phase diagram is given in the Fig.4(b), where the $FM$ state appears, the other states were already presented in the Fig.4(a).
\begin{figure}[h]
\begin{center}
\subfigure[ref1][]{\includegraphics[width=7.0cm,height=8.7cm,angle=-90]{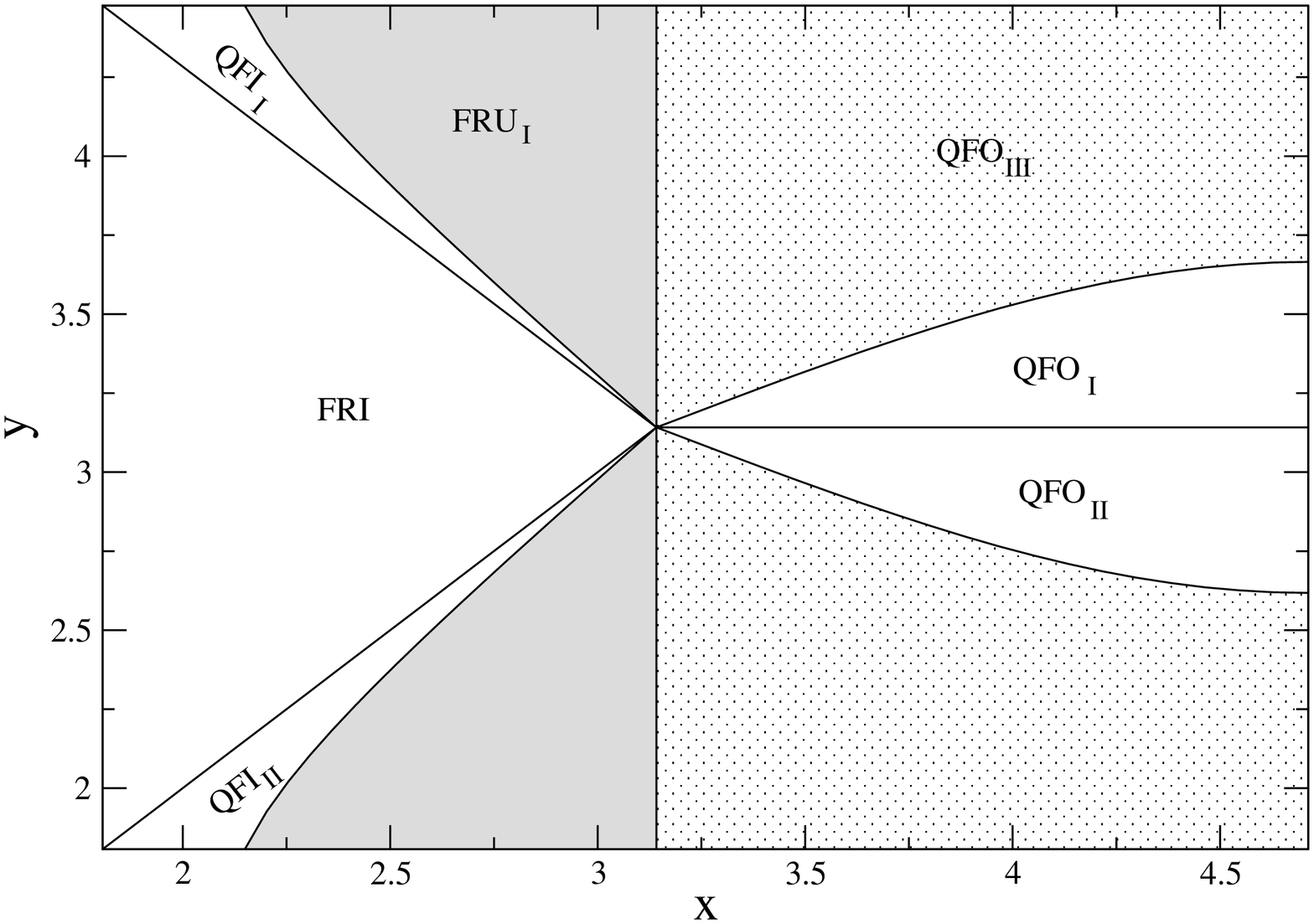}}
%\qquad
\subfigure[ref1][]{\includegraphics[width=7.0cm,height=8.7cm,angle=-90]{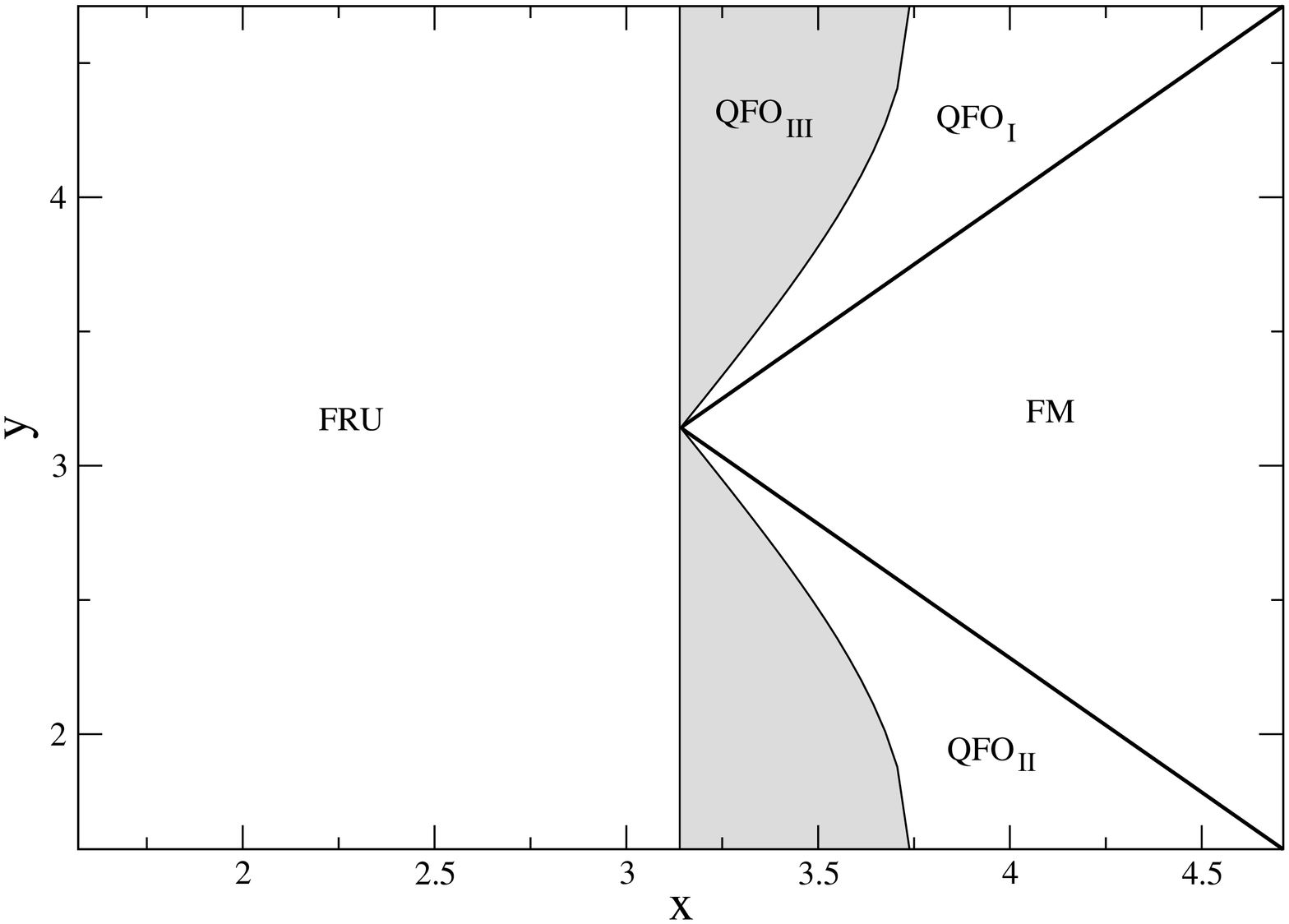}}
\end{center}
\caption{Phase diagram for the ATIH with spin-(1/2,1) chain. In (a) we have obtained it, in more general case, seven phase  diagrams, using the parameters given by eq.\eqref{sp4b} .  In (b) we have obtained it, restricting the interaction parameters to $J=J_{z}=\sin(x),J_{x}=J_{y}=2\sin(y)$, given by eq.\eqref{sp13}.}
\end{figure}

Finally it is quite interesting to mention that the state given by eq.\eqref{sp5} also appear when the next nearest interaction parameter considered is null, such as considered in reference \cite{Canova} i.e. when $\gamma=0$, and the restriction is extended over the decorated interaction parameters, for example, if we put 
\begin{equation}
J=\dfrac{1}{5}\sin(x), \text{ \ \ } J_{x}=J_{y}=J_{z}=4\sin(y), \text{ \ \ } \gamma=0,\label{sp14}
\end{equation}
we obtain the more simple phase diagram with four states as displayed in the Fig.5. The $FM$ and the $FRI$ states are separated by the $x=\pi$ line and they end up in the $QFO_{III}$ and $FRU_{I}$. Similar phase diagram was presented in the work \cite{Canova} with zero external magnetic field and zero next-nearest interaction parameter. In the work \cite{Canova} they also consider an external magnetic field but with zero next nearest interaction parameter where a similar phase diagrams were obtained, as displayed in the Fig.4 and 5(a). So we conclude that the inclusion of the ($s_{c},s_{c}$) interaction enable us to investigate a rich number of states even in the absence of an external magnetic field. Both restrictions (\ref{sp13}-\ref{sp14}) give the value for the ground-state energy $E_{0}=0$.

\begin{figure}[h]
\begin{center}
\includegraphics[width=7.cm,height=8.7cm,angle=-90]{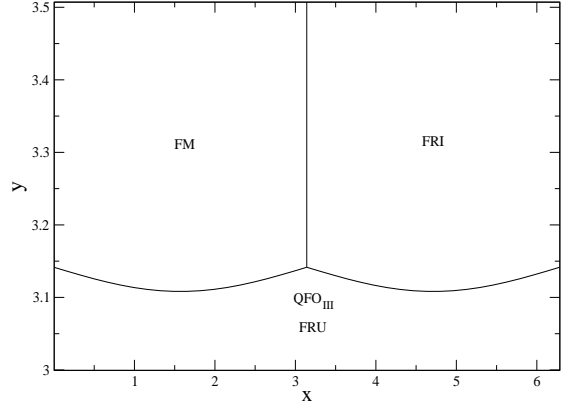}%
\end{center}
\caption{Schematic representation, for the connectors with spin-$1/2$ and decorated Heisenberg spin-$1$, of the phase diagram for the decorated $XXZ$ spin-1 model showing the presence of the $FM$ state. We fixed the parameter interaction $J_{x}=J_{y}=J_{z}=4\sin(y),J=1/5\sin(x) $ and $\gamma=0$, obtaining only three states. The $FM$, $FRI$ are non-degenerated while the $QFO$ and $FRU$ are two-fold degenerated. These sectors converge into one tricritical point $(\pi,\pi)$.}
\end{figure}

\section{The ATIH chain thermodynamics}

Thermodynamics properties could be studied using the known decorated transformation spin proposed in reference \cite{Fisher,Domb}.  Let us write the partition function as follow
\begin{equation}
{\mathcal Z}=\sum\limits_{\left\{ s_c\right\} }\prod\limits_{i=1}^{N}\text{Tr}_{\left\{S\right\} }{\rm e}^{-\beta H_{i,i+1}}=\sum\limits_{\left\{ s_c\right\}}\prod\limits_{i=1}^{N}w(s_c,s_c'),  \label{d1}
\end{equation}%
where $N$ is the number of decorated bounds,
whereas the Tr$_{\{S\}}$\ stands for the trace of the central
decorated system or Heisenberg {\it interaction edge} (in our case), while by $w(s_c,s_c')$ we represent the Boltzmann weight.
One should notice that the transformation (\ref{d1}) is rather general, since
it is valid for arbitrary spin values contained in the decorated site. The
set $\{ s_c\} $ represents the Ising {\it interaction vertex} and would take any spin value too. In this section we study the Ising-Heisenberg chain where the Ising {\it interaction vertex} take the spin values $S=1/2$ or 1.

 The ATIH chain partition function can be expressed as 
\begin{equation}
{\mathcal Z}=f^{N}{\mathcal Z}_{0},  \label{d3a}
\end{equation}%
where ${\mathcal Z}_{0}$ is the partition function of the effective Ising chain with arbitrary spin-$S$, whereas $f$ means a constant for the effective Ising chain.

When decorated Heisenberg {\it interaction edge} sites are occupied by spin-1/2 or 1, it is necessary to perform the partial trace over all those decorated spin sites. 

\subsection{The Ising {\it interaction vertex} with spin-$1/2$}
In order to map the ATIH chain into an effective Ising chain,
let us consider the Ising  {\it interaction vertex} with spin-1/2. For this case the associated Boltzmann weight function $w(s_c,s'_c)$ has the form
\begin{equation}
w(s_c,s'_c) ={\rm e}^{\lambda _{+}^{(1)}}+{\rm e}^{\lambda
_{+}^{(2)}}+{\rm e}^{\lambda _{-}^{(2)}}+{\rm e}^{\lambda _{-}^{(1)}}, \label{d4}
\end{equation}%

where $\lambda _{\pm}^{(1)}$ and $\lambda _{\pm}^{(2)}$ are given by eq.\eqref{a5} for $S=1/2$ (spin of decorated sites) whereas the associated Boltzmann weight for spin $S=1$ could be obtained using the eqs.\eqref{a8-1}-\eqref{a8}, then the ATIH chain model considered here is written by

%\begin{equation}
%w(s_c,s'_c) = 2{\rm e}^{-\beta(\gamma +\frac{1}{4}J_z)}\cosh(\beta\frac{1}{4}\sqrt{16\alpha^2+J_{-}^2})+2{\rm e}^{-\beta(\gamma-\frac{1}{4}J_z)}\cosh(\beta \frac{1}{4}J_+) \label{d4-1}
%\end{equation}%

\begin{eqnarray}
w(s_c,s'_c) = \left\{
\begin{array}{l}
2{\rm e}^{-\beta(\gamma +\tfrac{1}{4}J_z)}\cosh\big(\beta\tfrac{1}{4}\sqrt{16\alpha^2+J_{-}^2}\big)+2{\rm e}^{-\beta(\gamma-\frac{1}{4}J_z)}\cosh(\beta \frac{1}{4}J_+),\quad S=1/2 \\
{\rm e}^{-\beta\gamma}\big({\rm e}^{\beta J_z}+2{\rm e}^{\tfrac{\beta}{2}J_z}\cosh\big(\tfrac{\beta}{2}\sqrt{J_z^2+2J_{+}^2}\big)+4\cosh(\tfrac{\beta}{2} J_{+})\cosh(\beta\alpha)+2{\rm e}^{-\beta J_z}\cosh(\beta 2\alpha)\big),\quad S=1\\
\end{array}
\right.\label{d4-1}
\end{eqnarray}

The effective Ising chain partition function is
represented by their Boltzmann weight function $\widetilde{w}(s_c,s'_c)$ which read as
\begin{equation}
\widetilde{w}\left( s_c,s_c'\right) =f\exp \left\lbrace -\beta \left( Ks_{c}s'_{c}+B(s_{c}+s'_{c})\right)\right\rbrace,   \label{d2}
\end{equation}%

Using decorated transformation, we obtain the new parameters for the effective Ising chain,

\begin{align}
f^{2}=&w(\tfrac{1}{2},\tfrac{-1}{2})\sqrt{w(\tfrac{1}{2},\tfrac{1}{2})w(\tfrac{-1}{2},\tfrac{-1}{2})},\notag\\
 -\beta K=&4\ln \left( \frac{w(\tfrac{1}{2},\tfrac{1}{2})w(\tfrac{-1}{2},\tfrac{-1}{2})}{w(\tfrac{1}{2},\tfrac{-1}{2})^2}\right) ,\notag\\
-\beta B=&\frac{1}{2}\ln \left( \frac{w(\tfrac{1}{2},\tfrac{1}{2})}{w(\tfrac{-1}{2},\tfrac{-1}{2})}\right),
\label{d6}
\end{align}%
where the new effective parameters of Ising chain can be expressed as a function of the parameter of the original Hamiltonian. Thus $f$ is just a constant, whereas $K$ means a coupling parameter and finally $B$ corresponds to the external magnetic field.

The expression for the partition function  of ATIH chain results in 
\begin{equation}
{\mathcal Z}=f^{N}{\mathcal Z}_{0}=f^{N}\sum\limits_{\{s_c \}}\prod\limits_{i}^{N}{\rm e}^{-\beta
\left( K s_{c}s'_{c}+B(s_{c}+s'_{c})\right) }.  \label{d7}
\end{equation}

Using the eq.\eqref{d7} we are able to map the asymmetric tetrahedral spin-($1/2,S$) Ising-Heisenberg chain, into an effective spin-1/2 Ising chain. Where the spin $S$ of Heisenberg {\it interaction edge} could be 1/2 or 1.

\subsection{The Ising {\it interaction vertex} with spin-$1$}

Another case that we consider will be the Ising {\it interaction vertex} with spin-1. Thus the asymmetric tetrahedral spin(1,S) Ising-Heisenberg chain, will be mapped into an effective spin-1 chain. Similar to the previous case  we obtain the following Boltzmann weight expressed as follow
\begin{eqnarray}
{\widetilde w}( s_c,s'_c) &=&f\exp \left\lbrace -\beta ( K_{1}{s_c}{s'_c}+B({s_c}+{s'_c})+D({s_{c}}^{2}+{s_{c}}^{2})\right. 
\left.+E({s_{c}}^{2}s'_{c}+s_{c}{s'_c}^{2})+K_{2}{s_c}^{2}{s'_c}^{2})\right\rbrace ,
\label{d3}
\end{eqnarray}%
where $K_1$, $K_2$, $B$, $D$ and $E$ are the parameter to be determined.

The new parameters of eq.\eqref{d3} can be expressed using the associated Boltzmann weight, which is written as
\begin{align}
f =&w(0,0),\label{sp1-tr-1}\\
-\beta K_{1}=&\frac{1}{4}\ln \left( \frac{w(1,1) w(-1,-1) }{w(0,0)^{2}}\right) ,\text{ \ }\label{sp1-tr-2}\\
-\beta B=&\frac{1}{2}\ln \left( \frac{w(1,0) }{%
w( -1,0) }\right) ,  \label{sp1-tr-3} \\
-\beta D=&\frac{1}{2}\ln \left( \frac{%
w( 1,0) w\left( -1,0\right) }{w(0,0)^{2}}\right) ,\label{sp1-tr-4}\\ 
-\beta E =&\frac{1}{4}\ln \left( \frac{w\left( 1,1\right)
w(1,0)^{2} }{w\left( -1,-1\right) w(-1,0)^{2} }%
\right),\label{sp1-tr-5} \\  
-\beta K_{2} =&\frac{1}{4}\ln \left( \frac{w\left(1,1,\right)
w\left(-1,-1\right) w(0,0)^{2}}{w\left(1,0\right)^2 w\left( -1,0\right)^2 }\right). \label{sp1-tr-6} 
\end{align}%

Similar to the previous case $f$ means just a constant in the new effective 
Hamiltonian, while $K_1$ being the coupling parameter, whereas B corresponds to the external magnetic field, the parameter $D$ represents the single-ion anisotropy, whereas $E$ correspond to the interaction of quadratic and linear interaction among the nearest spin and finally $K_2$ are the parameter of the biquadratic interaction.

In this case if we consider a null magnetic field, we have  $
w(1,1)=w(-1,-1)$ and $w(1,0)=w(-1,0)$.  Under this condition the eqs.\eqref{sp1-tr-3} and eq.\eqref{sp1-tr-5} leads to $B=0$ and $E=0$, respectively, then the  Boltzmann weight function $w(s_c,s_c')$ defined by eq.\eqref{d3}, reduce the following relation%
\begin{equation}
{\widetilde w}( s_c,s'_c) =f\exp \Big( -\beta \big(K_{1}s_{c}s'_{c}+D ({s_c}^{2}+{s'_c}^{2})+K_{2}{s_c}^{2}{s'_c}^{2}\big)\Big),
\label{d9}
\end{equation}%

Finally we have concluded that, our mapping of the asymmetric tetrahedral spin-(1,S) Ising-Heisenberg chain can be expressed as an effective spin-1 Ising chain. Where as before the spin $S$ of Heisenberg {\it interaction edge} could be 1/2 or 1.

\subsection{The ATIH chain correlation functions}

We can notice that the partition function of the ATIH chain  obtained above, by mapping into Ising spin chain is limited. We cannot obtain directly the correlation function because the mapped Ising chain does not depend of the decorated spin. Then we can use the method presented by Fisher\cite{Fisher}, where the correlation function for the ATIH chain can be obtain using the decoration transformation in a similar way as was performed the partition function, assuming we known the correlation function of the effective spin-1 Ising chain.
 
Using the definition given in reference \cite{Fisher}, we have%
\begin{equation}
\left\langle S_{i}s_{k_1}s_{k_2}...\right\rangle =\frac{1}{\mathcal Z}%
\sum_{\{s_{k_j}\}}\sum_{S_i}S_{i}s_{k_1}s_{k_2}\dots{\rm e}^{-\beta H},  \label{co1}
\end{equation}%
here $S_{i}$ represents the decorated spin at site $i$, and $s_{k_j}$ are any spins of the systems along the chain, it could be either decorated spins $S$ or undecorated spins $s_c$. With ${\mathcal Z}$ as the total partition function of the
system and $H$ the total Hamiltonian. It is possible to split the above
relation in two parts, one of this being independent on the spin $S_{i}$ and
the other one  containing the $S_{i}$ dependence. Thus we write down the total Hamiltonian \eqref{a3} as 
\begin{equation}
H=H_{0}\left( S_{i},s_{c,1},s_{c,2},...\right) +H_{n}\left(
s_{c,1},s_{c,2},...\right) \equiv H_{0}+H_{n},  \label{co2}
\end{equation}%
where the $H_{0}$\ contains the dependence of the spins $(S_{i},s_{c,1},s_{c,2},...)$ and $H_{n}$ contains only combinations of the spins $\left( s_{c,1},s_{c,2},...\right) $. Actually, in the case when $\left[
H_{0},H_{n}\right] =0,$ we have
\begin{equation}
\left\langle S_{i}s_{k_1}s_{k_2}...\right\rangle =\frac{1}{\mathcal Z}%
\sum_{\{s_{k_j}\}}s_{k_1}s_{k_2}...{\rm e}^{-\beta H_{n}}\Omega(s_{c,i}s_{c,i+1}),  \label{co3}
\end{equation}%
and as was pointed out by Fisher\cite{Fisher}, it is possible to prove that $\Omega(s_{c,i}s_{c,i+1})$ can be represented as
\begin{equation}
\Omega(s_{c,i}s_{c,i+1})=\sum_{S_{i}}S_{i}{\rm e}^{-\beta H_{0}}. \label{co4}
\end{equation}%

We give some values for the correlation functions of the asymmetric tetrahedral Ising-Heisenberg
model in the case when the Ising {\it interaction vertex} with spins are equal to $s_c=1/2$ or $s_c=1$ and the decorated spin could be equal to $1/2$ or $1$. In the case of spin $s_c=1/2$, it is possible to obtain an equivalent form\cite{Fisher} for the right hand side of (\ref{co4}),
\begin{align}
\Omega(s_{c,i},s_{c,j}) = &\Big( q_{0}+q_{0,1}\big(s_{c,i}+s_{c,j}\big)+ q_{1,1}s_{c,i}s_{c,j}\Big) \sum_{S_{i}}{\rm e}^{-\beta H_{0}}.
\end{align}

As an example, let us apply to evaluate the following correlation, with arbitrary sites $i$ and $j$, instead of performing only among next nearest sites\cite{Canova}, considering $S_{a}^z$ could be spin 1/2 or 1, thus the correlation read as
\begin{align}\label{corr-prop}
\langle S_{a,i}^{z}s_{c,j} \rangle =q_{0}\langle s_{c,j} \rangle +q_{0,1}\big( \langle s_{c,i}s_{c,j} \rangle + \langle s_{c,i+1}s_{c,j} \rangle \big) + q_{1,1} \langle s_{c,i}s_{c,i+1}s_{c,j}\rangle
\end{align}
where the coefficients $q$'s can be obtained solving the system equation \eqref{co3} and \eqref{corr-prop}, from where we verify their solution is given as a derivative of the parameters obtained in \eqref{d6} respect to magnetic field $h$, which read as
\begin{align}
q_{0} = -\dfrac{1}{2\beta}\dfrac{\partial}{\partial h} \ln f, \qquad
q_{1,0} = \dfrac{1}{2} \dfrac{\partial B}{\partial h},\qquad\text{and}\quad
q_{1,1} = \dfrac{1}{8}\dfrac{\partial K}{\partial h},
\end{align}
then it is possible to write as a combination of the correlation function of the effective Ising chain with up to three-body spin correlations, this correlation is given explicitly as follow
\begin{align}
\langle s_{c,i}s_{c,j} \rangle  = &\langle s_c\rangle^2 + \big(1-\langle s_c\rangle ^2 \big) {\rm e}^{(i-j)/\xi}   \\
\langle s_{c,i}s_{c,i+1}s_{c,j} \rangle =&\langle s_c\rangle^3 + \langle s_c\rangle \big(1-\langle s_c\rangle ^2 \big) \Big({\rm e}^{(i-j)/\xi}+{\rm e}^{-1/\xi}(1 +{\rm e}^{(j-i)/\xi}) \Big)
\end{align}
where $\xi$ is the correlation length of the effective(standard) Ising chain\cite{baxter}.

The auto correlation function of the $\langle (S_a^z)^2\rangle$ when we consider the Ising {\it interaction vertex} $s_c=1/2$, simply becomes a constant equal to $1/4$, but when we consider spin-1, this expression becomes non-trivial and we can obtain analogous to previous correlation discussed, then the general expression can be write as
\begin{align}
\left\langle \left( S_{a,i}^{z} \right)^{2}  \right\rangle= Q_{0} +2Q_{1,0} \left\langle s_{c,i} \right\rangle  + Q_{1,1} \left\langle s_{c,i},s_{c,i+1} \right\rangle
\end{align}
onces again the coefficients $Q$'s can be obtained using the following relation
\begin{align}
Q_{0} =& \dfrac{1}{4} \left[ M(\tfrac{1}{2},\tfrac{1}{2})+ M(-\tfrac{1}{2},-\tfrac{1}{2})+  2M(\tfrac{1}{2},-\tfrac{1}{2}) \right], \\
Q_{1,0} =& \tfrac{1}{2} \left[ M(\tfrac{1}{2},\tfrac{1}{2})- M(-\tfrac{1}{2},-\tfrac{1}{2}) \right],
\\
Q_{1,1} =& M(\tfrac{1}{2},\tfrac{1}{2})+ M(-\tfrac{1}{2},-\tfrac{1}{2})-  2M(\tfrac{1}{2},-\tfrac{1}{2}),
\end{align}
with
\begin{equation}
M(s_{c,i},s_{c,i+1})=\dfrac{1}{\beta} \dfrac{\partial w(s_{c,i},s_{c,i+1})}{\partial J_{z}}+\dfrac{1}{2\beta ^2} \dfrac{\partial^2 w(s_{c,i},s_{c,i+1})}{\partial h^2},
\end{equation}
whereas $w(s_{c,i},s_{c,i+1})$ already was defined in eq.\eqref{d4-1}.
Other correlation functions like $\langle S_{a,i}^{\nu}S_{b,i}^{\nu'}\rangle$, are null when $\nu\ne \nu'$. Whereas for  $\nu=\nu'$,
the correlation function can be obtained directly using the derivatives of the free energy instead of using the previous iterative method such as the one performed by Canova\cite{Canova}, particularly we show the following nearest correlation,

\begin{eqnarray}
\left\langle S_{a}^{z}\right\rangle  &=&-\frac{1}{\beta}\frac{\partial\ln({\mathcal Z})}{\partial h}\\
\left\langle S_{a,i}^{\nu}S_{b,i}^{\nu}\right\rangle 
&=&-\frac{1}{\beta}\frac{\partial\ln({\mathcal Z})}{\partial J_{\nu}}, \quad \text{with} \quad \nu=\{x,y,z\}.
\end{eqnarray}%

We remark that above conclusion are also valid even for spin $s_c=1$.
Similar analysis could be performed to obtain the correlation function when $s_c=1$, in this case the $\Omega(s_{c,i},s_{c,i+1})$ function will be defined as
\begin{align}
\Omega (s_{c_{i}},s_{c,j}) = &\left\lbrace q_{0}+q_{0,1}\big(s_{c,i}+s_{c,j}\big)+ q_{1,1}s_{c,i}s_{c,j}+ q_{1,2}\big(s_{c,i}s_{c,j}^2 + s_{c,i}^2 s_{c,j}\big)+ q_{0,2}\big(s_{c,i}^2 + s_{c,j}^2\big)\right.\nonumber\\  &\left. + q_{2,2}s_{c,i}^2 s_{c,j}^2 \right\rbrace {\rm e}^{-\beta H_{0}},
\end{align}
using this relation, together with the correlation function of the effective Ising chain with spin-1, we can obtain other correlations functions for $s_c=1$, using the same recipes above. 

Alternatively we can obtain this kind of correlation using the direct transfer matrix formalism such as performed in reference \cite{Juan}, meanwhile the advantage of this method could be a non-iterative calculation. 
%\qquad 

%\newpage
\section{conclusions}
The phase diagrams of the asymmetric tetrahedral Ising-Heisenberg (ATIH) chain were studied for the case when the Ising {\it interaction vertex} is spin-$1/2$. Firstly we considered the XYZ {\it interaction edge} with spin-$1/2$,  and null  external magnetic field ($h=h_0=0$). The diagrams displayed in Fig. 2a, have shown seven states appearing in the model, with five critical transition points $(x,y)$ having three phase state converging and two critical transition points, where four phase states converge.  These states have shown their quantum nature for the XYZ {\it interaction edge} (decorated sites), for example, for the vector states (\ref{e3}) and  (\ref{e4}) we see that up and down orientation in the decorated sites have different probabilities defined by the factors $e_{1}(s_c,s'_c)$ and $e_{2}(s_c,s'_c)$. We also have analyzed the particular case when $J_{-}=0$, constructing the phase diagram and showing that $FM$ and $FRI$ states appear (Fig. 2b). Other situations have been also studied, particularly, the case where $\gamma=0,J_{x}=J_{y}$ and for the more simple configuration $\gamma=0,J_{x}=J_{y}=J_{z}$, (Fig. 3). Secondly, when the XXZ {\it interaction edge} (decorated) spin is 1, we have also obtained a rich phase diagram for the ground-state energy, even when an external magnetic field and the next nearest interaction are absent. These phase diagrams are shown in the figures 4 and 5. The $x=\pi$ line appears to be the limit of the $QFO$ and the other states, so for the critical point ($\pi,\pi$) the energy takes the value $E_{0}=0$. Some particular case of our results obtained have been compared with those obtained in reference \cite{Canova}.

 We have noticed  that using a decorated Ising model mapping transformation, initially given by Fisher \cite{Fisher}, the calculation of the partition function for the ATIH chain is reduced to a closed expression of Ising spin chain. We have considered some particular cases to discuss the thermodynamic properties, such as Ising {\it interaction vertex} with spin 1/2 and 1, whereas the {interaction edge} could be XYZ with spin-1/2 and XXZ with spin-1 respectively. The results for correlation function are presented generally for the situation when we have the Ising {\it interaction vertex} with spin-$1/2$ or 1 and the Heisenberg {\it interaction edge} with spins $1/2$ or $1$.  We have observed that some correlation function could be obtained using the derivative related to some parameter instead of using the decorated transformation used in reference\cite{Canova}, we also considered a long range correlation function where we used the decorated transformation to obtain the result.

{\centerline \textbf{\large Acknowledgments}} \vskip 0.5cm J.S.V. thanks FAPEMIG for full financial support, O. R. and S.M. de S. thank CNPq and FAPEMIG for partial financial support.

\end{document}